# A potential demand model for a multi-circulation feeder network design

Saeed Sani[1], Mehdi Ghatee[1]

[1]*Department of Mathematics and Computer Science, Amirkabir University of Technology, Tehran, Iran*
Coredponding Author e-mail: saeedsani@aut.ac.ir

## Abstract

**Background:** The development of public transit network can enhance the efficiency of the system as well as raise interest to use the system. Feeder bus service fills the area that is far from the railway system; therefore, designing a feeder network in a gap area causes the expansion of the main transit network.

**Purpose:** To present a modified potential demand model for designing a multi-circulation feeder network.

**Materials and Methods:** In this study, the potential demand is defined based on the traffic demand of each section related to each link, the average walking distance of passengers to specified links, the distance from the links to stations, and finally potential demand of accessing each main station for each link. A labelling method is used to create continuous circular and separated routes. After generating an initial solution using a constructive heuristic algorithm, a genetic algorithm was operated to solve the problem. In the second algorithm, each route is considered a gene, and each network is considered a chromosome. In fact, in this step, different routes from diverse solutions are combined as a unique feeder network. Tehran District 10 was selected as the case study and the model was tested on this area which is located near the central business districts. A machine learning approach has been applied to estimate the missing values of the related database.

**Results:** By running the algorithm, four feeder routes were obtained with reasonable travel times and distances of 4.8, 5.5, 5.8, and 8.1 km, starting and ending at the same railway station. Moreover, 98% of the area was covered by resulted feeder network with a maximum access distance of 300m.

**Conclusion:** What stands out from this method is that the combination of the modified feeder route design model and algorithms is feasible.

**Index Terms**: Operation research, Optimization, Public network transportation, Feeder network design

## 1. Introduction

Over the last century, urban transit network has noticeably developed in the world, particularly in big cities. In fact, many societies have faced an urban sprawl and inefficiency of public transit which caused some issues, such as high transit times, the tendency to use private cars and taxis, air pollution, increasing stress-related as well as respiratory-related diseases. So, to sort out these problems, experts and researchers are led to make public transit networks (PTN) more connected and easily accessible to attract more commuters to this network. One of the most significant characteristics of attracting more users is the accessibility to PTN which can potentially be considered for vulnerable people, especially elderly ones. Covering gaps by expanding the feeder network can propel us to a more efficient PTN. In other words, the feeder network could fill both residential areas and business districts and reduced walking distances and travel times. In many cities, underground railways and BRT are integrated to form an urban transit system. Both of them include stations which are the base of designing feeder routes. Then, choosing which station is more effective to allocate to a feeder route is a considerable aspect of cutting down the average travel time. In this paper, we use the term feeder bus and stop to make difference with vehicles and stations in the main network. Stops locating is resulted by population density and coverage of sections in a similar way as stations. Besides saving more time, feeder routes are more likely to meet the travel demand of residents..

The potential demand for feeder bus services, the distance between road and rail transit and the repetition factor of road bus lines are considered to establish a potential demand model for feeder route designing. Then the model was tested in a selected area in Shanghai (Zhu et al. 2017).

Ceder has designed a model with consideration to population density rate and maximum access time constraints of one feeder route to the system with an objective function of access rate (Ceder 2016). Kuah and Perl presented a heuristic model for designing a feeder-bus network to access an existing rail system which is considered under two different demand patterns, many-to-one (M-to-1) and many-to-many (M-to-M) (Kuah and Perl 1989). Martins has reported computing solutions for Feeder Bus Network Design Problem. The model which is a non-linear and non-convex mixed integer problem is aimed at minimizing a cost function including both passenger and operator interests (Martins and Pato 1998). Sze nee has represented a model for designing feeder routes with demand rate, frequency in requirement services and constraint length of each route with the objective function of passengers' costs and operator's cost (NEE 2004). Chandran et al take overall distances of travel in objective function in a VRP model with paying attention to capacity (Chandran and Raghavan 2008). Chien has represented an algorithm for designing feeder routes with consideration to demand rate, headways and services frequency. The target of this algorithm is to minimize the overall cost (I-Jy Chien 2005). Zhi-Chun Li et al have presented analytical models for optimizing a bus and rail transit system with feeder bus services by aiming optimization the frequencies and fares of the bus, rail and feeder bus services (Li, Lam and Wong 2009). Xiugang Li et al have developed an analytical model by assuming high demands and long length of the service area and a two-vehicle operation for each zone (Li and Quadrifoglio 2011). Martínez and Eiró have designed a routing model for feeder services in which the demands of entrance and exit scheduling to each stop and costs are considered in the objective function (Martínez and Eiró 2012). Hu et al analyzed the characteristics of urban rail transit and conventional buses. Passengers' distribution time, travel distance, walking time, waiting time and operational speeds have been assumed to construct a model for the layout region of feeder buses (Hu, Zhang and Wang 2012).

Saadia Tabassuma used the Gini coefficient and other methods to evaluate the existing feeder modes available for BRT System. Route reliability, travel time, travel cost, comfort, safety and security environmental aspects are considered as main factors in calculating Travel impedance (Tabassum et al. 2017). Chandra and Quadrifoglio used a new street connectivity indicator to predict transit performance. The role of street network connectivity in service quality level has been identified for prediction (Chandra and Quadrifoglio 2013). Park et al have represented an optimization approach to designing feeder bus routes based on a neural network-based embedding methodology which is applied to real-time taxi GPS data. This approach is aimed to contribute to a reduction in traffic during rush hours (Park, Lee and Sohn 2019). Guo et al represented a bi-objective optimization model which considered the maximization of travel demand, as well as the minimization of travel distance; then, the exact-constraint method is used to generate Pareto optimal solutions (Guo et al. 2019). Deng et al have considered minimization of passenger travel cost and transit operating cost to represent a model for designing the feeder-bus network. Passenger travel cost includes bus waiting cost, accurate bus riding cost, transfer cost between buses, trains and rail riding cost. Also, a new genetic algorithm for determining optimal feeder-bus operating frequency has been presented (Deng et al. 2014). CHEN et al proposed a robust model by mixing driver's mode choice as well as route choice based on a combination of user equilibrium and cross nest logit. Travel time uncertainty on roads and multi-class demands is the main contributions which are considered in this research (Chen et al. 2017). Almasi et al proposed a strategy for designing transit networks. They have used an integrated multi-modal transit model in order to design a better combination of different mode transit systems (Almasi et al. 2018).

Lee et al developed an algorithm for optimal flexible feeder bus routing by considering the relocation of buses for multiple stations and trains, minimizing the total cost and limiting the maximum degree of circuity for each passenger (Lee et al. 2019). Ramos et al presented a model to solve the waste collection routing problem. This model takes into account the maximization of waste collection along with the minimization of transport costs (Ramos, de Morais and Barbosa-Póvoa 2018). Huang et al represented a new network design optimization model for the responsive customized bus (CB) problem. The model is two-phased: dynamic and static by considering inserting passenger requests dynamically in an interactive manner and optimizing the service network statically based on the overall demand respectively (Huang et al. 2020).

**Table1. Characteristics of obtained feeder routes.**

In many cases, researchers have concentrated on designing one feeder route which is connected to the main route station. So, covering a vast area by more than one feeder route is another problem that cannot be met by these kinds of models. In this paper, we introduce a model for designing feeder routes in an area with more accessible PTN stations. In fact, the model is capable of optionally choosing the more effective stations and allocating appropriate feeder routes to them. This model is the expansion of a route design model of feeder bus service which used potential demand values to optimize the problem (Zhu et al. 2017). By using three-dimensional variables -two indexes for specifying connected stops and one index for routes- and new connection constraints -guaranteeing the connectivity of each route- we can design a multi-routes feeder network. In addition, not only have we altered the structure of Zhu's model to make the model capable of designing more routes, but we set a new method to calculate potential demand values which is more efficient than the last one.

## 2. Mathematical Model

### 2.1. Model Construction

In this paper, we use potential demand values to strengthen links which are candidates for being a part of feeder routes. Traffic demand of the area straightly and the average walking distance of commuters inversely both affect the measure of potential demand value for each link. The Significance of connecting a link to each PTN station is another influential value, so the distance from a link to the main network station and the demand for accessing the main network station have noticeable impacts on potential demand. Considering these characteristics, the potential demand for a link to be used in a feeder route –allocating to the PTN station $s$- can be calculated as

$$Pd_{ijs} = \frac{Td_{ij}}{S_{ij}} L_{ijs} e^{g_{ijs}}$$

$$g_{ijs} = \begin{cases} 1 & if\ d_{ijs} \leq \frac{1}{3}\max_S d_{ijs} + \frac{2}{3}\min_S d_{ijs} \\ 2 & if\ \frac{1}{3}\max_S d_{ijs} + \frac{2}{3}\min_S d_{ijs} < d_{ijs} \leq \frac{2}{3}\max_S d_{ijs} + \frac{1}{3}\min_S d_{ijs} \\ 3 & if\ d_{ijs} > \frac{2}{3}\max_S d_{ijs} + \frac{1}{3}\min_S d_{ijs} \end{cases}$$

$Pd_{ijs}$ is the potential demand of link $(i,j)$ operating feeder route $s$ which is allocated to PTN station $s$.

$Td_{ij}$ is the traffic demand of link $(i,j)$.

$S_{ij}$ is the average walking distance of passengers to link $(i,j)$.

$L_{ij}$ is the distance from the link $(i,j)$ to the PTN station $s$.

$d_{ijs}$ is the demand for reaching PTN station $s$ from link $(i,j)$.

$g_{ijs}$ specifies the level of demand from link $(i,j)$ for reaching PTN station $s$.

In order to calculating $g_{ijs}$, first, for each $(i,j)$ the set of $\{d_{ijs}|s \in S\}$ is divided into three portions based on the quantities of members. Then, we set 1, 2 and 3 to lower, middle and upper level respectively. Using the exponential function is due to significance of reducing the whole travel times in the expanded network.

Zhu's model designs a one-route feeder network by potential demand values, in which internal characteristics are considered for the calculation of these values. However, our model designs a multi-route network with taking into account users' destinations in the main network to calculating potential demand values (some areas may conceivably require more than one route to be covered efficiently and allocating each route to more appropriate station plays an integral role). In this paper, it is presumed the same assumption about dividing the traffic zone as the mentioned research (Zhu et al. 2017).

The graph $G = (N, A)$ shows the set of links and nodes in the study area (Ceder 2016). We assign $N = N' \cup S$ in which S is the set of all PTN stations and $N'$ is the set of other nodes within the area. Depending on the study area, one or more circular feeder routes will be resulted from solving the model. Every resulted feeder route starts from a PTN station and ends at it. Labelling each node helps the model to obtain unique, circular feeder routes. The model maximizes the potential demand of commuters, satisfying the constraints of allocating each link to a PTN station, maximum travel distance and making a closed loop. Based on these needs, the model can be achieved as

$$Max \sum_{(i,j)\in A} \sum_{s\in S} Pd_{ijs} \times x_{ijs} \quad (1)$$

$$subject\ to: \sum_{i\in N'} \sum_{j\in S-\{s\}} x_{ijs} = 0 \quad \forall s \in S \quad (2)$$

$$\sum_{h\in N} x_{ihs} - \sum_{l\in N} x_{lis} = 0 \quad \forall i \in N, \forall s \in S \quad (3)$$

$$\sum_{i\in N'} \sum_{j\in S} \sum_{s\in S} x_{ijs} \leq M \quad (4)$$

$$\sum_{i\in N} \sum_{j\in N} t_{ij} \times x_{ijs} \leq T \quad \forall s \in S \quad (5)$$

$$x_{ijs} \times (b_j - b_i - 1) = 0 \quad \forall i \in N, j \in N', s \in S \quad (6)$$

$$b_i = 0 \quad \forall i \in S \quad (7)$$

$$x_{ijs} \in \{0,1\} \quad \forall (i,j) \in A, \forall s \in S \quad (8)$$

$$0 \leq b_i \leq n(N) \quad \forall i \in N \quad (9)$$

Constraint (2) guarantees that the feeder route $s$ should be allocated to the main network station $s$. Equation (3) is constraint of creating a circular route in the network. Constraint (4) represents the constraint of the maximum number of feeder lines. Constraint (5) limits the maximum travel time. Equations (6) and (7) are constraints of labelling all nodes of the feeder routes which guarantee that the feeder routes are connected to the main network stations. If a feeder route passes from a node, the main network station also joins it. Constraints (8) and (9) also specify the type of decision variables. Equation (8) is the definition of $x_{ijs}$, where $x_{ijs} = 1$ indicates that link $(i,j)$ is a part of the feeder route $s$ and otherwise $x_{ijs} = 0$.

In this new model, potential demand is considered as a main value which adverts users' destinations, and this is more likely to make more efficient connections between stops and stations. Furthermore, by using special constraints, this model can design a multi-route network simultaneously in order to form a more effective feeder network. Heuristic and meta-heuristic algorithms are appropriate to solve the model.

Solving this NLP model results in, depending to area, one or several closed, circular feeder routes connecting to the railway network. Besides decreasing access time to the main network, reducing the operating cost is considered in the model. Moreover, reduction of travel time through the network is attended in model by new calculation of potential demand.

In addition, we can transform this model into a LP model. It is possible by using (10) and (11) instead of equation (6).

**Theorem.** *Following constraints function as constraint (6) in the proposed model, where $\mathcal{B}$ is a big number.*

$$\mathcal{B} \times x_{ijs} + b_j - b_i - 1 \leq \mathcal{B} \quad \forall i \in N, j \in N', s \in S \quad (10)$$

$$\mathcal{B} \times x_{ijs} - (b_j - b_i - 1) \leq \mathcal{B} \quad \forall i \in N, j \in N', s \in S \quad (11)$$

Proof. $\forall i \in N, j \in N', s \in S$, if $x_{ijs} \times (b_j - b_i - 1) = 0$, then $x_{ijs} = 0$ or $(b_j - b_i - 1) = 0$. There are three possibilities:

I) $x_{ijs} = 0, (b_j - b_i - 1) \neq 0$.

II) $(b_j - b_i - 1) = 0, x_{ijs} = 1$.

III) $x_{ijs} = 0, (b_j - b_i - 1) = 0$.

Let $x_{ijs} = 0$ and $(b_j - b_i - 1) \neq 0$ then $\mathcal{B} \times x_{ijs} + b_j - b_i - 1 = b_j - b_i - 1 \leq \mathcal{B}$, and $\mathcal{B} \times x_{ijs} - (b_j - b_i - 1) = -(b_j - b_i - 1) \leq \mathcal{B}$.

Also, Let $(b_j - b_i - 1) = 0$ and $x_{ijs} = 1$ then $\mathcal{B} \times x_{ijs} + b_j - b_i - 1 = \mathcal{B} \leq \mathcal{B}$, and $\mathcal{B} \times x_{ijs} - (b_j - b_i - 1) = \mathcal{B} \leq \mathcal{B}$.

Moreover if $x_{ijs} = 0$ and $(b_j - b_i - 1) = 0$ then $\mathcal{B} \times x_{ijs} + b_j - b_i - 1 = 0 \leq \mathcal{B}$, and $\mathcal{B} \times x_{ijs} - (b_j - b_i - 1) = 0 \leq \mathcal{B}$.

Thus $\forall i \in N, j \in N', s \in S$ which $x_{ijs} \times (b_j - b_i - 1) = 0$, (10) and (11) are correct.

On the other hand, $\forall i \in N, j \in N', s \in S$, if $\mathcal{B} \times x_{ijs} + b_j - b_i - 1 \leq \mathcal{B}$ and $\mathcal{B} \times x_{ijs} - (b_j - b_i - 1) \leq \mathcal{B}$, then

$$\begin{cases} \mathcal{B} \times x_{ijs} + b_j - b_i - 1 \leq \mathcal{B} \rightarrow b_j - b_i - 1 \leq \mathcal{B}(1 - x_{ijs}) \\ \mathcal{B} \times x_{ijs} - (b_j - b_i - 1) \leq \mathcal{B} \rightarrow -(b_j - b_i - 1) \leq \mathcal{B}(1 - x_{ijs}) \end{cases}$$

Obviously, there are two conditions:

I) $x_{ijs} = 0$.

II) $x_{ijs} = 1$.

Let $x_{ijs} = 0$ then $x_{ijs} \times (b_j - b_i - 1) = 0 \times (b_j - b_i - 1) = 0$.

We assume that $x_{ijs} = 1$, it follows that

$$\begin{cases} b_j - b_i - 1 \leq \mathcal{B}(1 - x_{ijs}) = 0 \\ -(b_j - b_i - 1) \leq \mathcal{B}(1 - x_{ijs}) = 0. \end{cases}$$

So,

$$\begin{cases} b_j - b_i - 1 \leq 0 \\ -(b_j - b_i - 1) \leq 0 \end{cases}$$

We obtain $b_j - b_i - 1 = 0$ therefore $x_{ijs} \times (b_j - b_i - 1) = x_{ijs} \times 0 = 0$.

Hence, $\forall i \in N, j \in N', s \in S$ which $\mathcal{B} \times x_{ijs} + b_j - b_i - 1 \leq \mathcal{B}$ and $\mathcal{B} \times x_{ijs} - (b_j - b_i - 1) \leq \mathcal{B}$, constraint (6) is correct.

## 2.2. Solution Algorithm

The represented feeder route design model is an NP-hard problem with one-dimensional and three-dimensional variables. Depending on the size of the study area, a large-scale problem may be needed to be generated. The number of variables and constraints are $(n(N))^2 n(S) + n(N')$ and $n(N)n(N')n(S) + n(N)n(S) + 3n(S) + 1$ respectively. For instance, a problem with 50 stops and 10 stations has 36050 variables and 30631 constraints. So, using some heuristic algorithms is a faster way to solve this proposed operation research problem. Using constructive heuristic algorithms for solving VRP and GVRP can produce some initial solutions (Pop et al. 2011). Then, by performing a genetic algorithm, with a multi-parent crossover which uses a diversity operator, obtaining a good answer can be achievable (Elsayed, Sarker and Essam 2014). In this step, we use a function to produce three offspring by three parents with the replacement of routes in parents:

$$o_1 = F(X_1, X_2, X_3)$$

$$o_2 = F(X_2, X_3, X_1)$$

$$o_3 = F(X_3, X_1, X_2)$$

where

$$X_i = \{r_s^i | s \in S, r_s^i = \{stop_1^i, stop_2^i, \ldots, stop_k^i | stop_k^i \in N'\}\}$$

Finding $F$ is as follows:

1: $X_f = X_i$

2: $For\ each\ s \in S\ if\ Length(r_s^{i'}) > Length(r_s^{i''})\ then\ r_s^f = r_s^{i'}$

3: $F(X_i, X_{i'}, X_{i''}) = X_f$

Also we perform a mutation step for each offspring by a improvement heuristic algorithm (Pop et al. 2011).

**Figure1. Flowchart for the proposed algorithm to solve the problem.**

After producing new solutions, they should be checked for the model's constraints satisfaction; subsequently, the best solution -with the highest amount of objective function- can be gained. Like other works, this model considers different crucial aspects such as travel demand, walking distance, travel time and route numbers to reduce the number of the required fleet. In this model, the priority is satisfying users to reach their destinations in a shorter time. Using a new mixed algorithm to solve the problem leads us to reach better solutions in a short time.

The model and solution algorithm have been implemented on a simple network with 21 stops, and 4 main network stations. After having generated 6 initial solutions, we produced 100 generations using the algorithm. Finally, the best solution has been selected with the biggest amount of objective function (3154120.5) and three feeder routes.

**Table2. Characteristics of obtained feeder routes.**

**Figure2. Obtained feeder routes for a simple network.**

3. Experimental Results

   3.1. Data Preparation

In this article, we used a statistical database of Tehran Municipality, and we aimed the Tehran district 10 as the case study. Calculation of the potential demands for modelling the data needs different characteristics' values which are not completely clear in the mentioned database. Tackling this problem and completing missing values, we used a novel machine learning tool and developed random forest and local least squares (DRFLLS). This method applies seven categories for estimation of the optimal number of missing values' neighbourhoods; then, the local least squares method is used to estimate the missing values (Al-Janabi and Alkaim 2020).

The traffic demand of each link, the average walking distance of passengers to each link, the distance from each link to each PTN station, the demand for reaching each PTN station from each link, population density in the specified neighbourhood of each link, the average age of people in the specified neighbourhood of each link, and the housing density in the specified neighbourhood of each link are used as the features which have been applied to estimating the missing values by DRFLLS. Clearly, applying more related features from an accessible database may conceivably make better approximations.

### 3.2. Method Performance and Results

By using MATLAB software in a computer (CPU :core i7, RAM: DDRIII 8GB), we have performed aforementioned algorithm to solve the proposed model for Tehran district 10; in addition, total time for solving was 90.28 second. This Genetic Algorithm by using 30 initial parents and generating 100000 generations found a unique solution (second offspring from generation 19956) with the biggest amount of objective function (1.9099e+07).

Moreover, we ran the algorithm on three other samples, with benchmark and randomized initial solutions, and compared them based on various aspects.

**Table3. Comparing results of four different samples solved by the algorithm.**

**Figure3. Resemblance between final answer and initial solutions for four samples.**

Figure 3 (drawn by specifying similar links between best answer and initial solutions, and calculating the centers of gravity) shows the amount of resemblance between best answers and initial parents in four samples. Also, the black circles in the big circle depict the best answers, and the black circles on the big circles' circumferences display initial solutions. The size of each black circle indicates the magnitude of its objective function.

These charts reveal the importance of initial solutions, as well as their impacts on finding appropriate final answers. In fact, they represent how much of the initial networks' structures have been used by the algorithm to form the final feeder networks.

### 4. Case Study

The proposed model is implemented in Tehran district 10 with 4 underground and 9 BRT stations at the north, east and south. This area is the house of about 327,000 people, making the area much more significant to connecting to public transit network. The test area was divided to grids with the size of 100m×100m.

By considering traffic demand in each link, calculating $S_{ij}$ by the formula Zhu used (Zhu et al. 2017), shortest path distances between links and stations and calculating $g_{ijs}$, the matrix of $Pd$ was obtained. $g_{ijs}$ is gained by finding $d_{ijs}$ calculated by the data of Origin-Destination travels statistics. We assume $Pd_{ijs} = Pd_{jis}$ to half the scale of calculation. $t_{ij}$ depends on the length of link $(i, j)$, traffic level between node $i$ and $j$ and the average speed and stopping time of the feeder bus. Maximum traveling time of each feeder bus has been set to $T = 45$ min. Also, maximum number of the feeder routes has been considered to $M = 5$ in this area.

The actual, accurate road network of Tehran District 10 has been used to test the suggested model. We used MATLAB software to solve the mathematical model by mentioned algorithms. The Final solution is a four-route network in lengths of 4.8, 5.5, 5.8, and 8.1 km. Figure 4 shows the obtained feeder network, opting one BRT and three underground railway stations as destinations. As it is clear in figure 5, the resulted feeder network has covered more than 98% of the area with maximum access distance of 300m. Another significant result is the rates of $g_{ijs}$ in following table.

**Table4. Three levels of satisfied demand from links for reaching PTN stations in final feeder network.**

**Figure4. Proposed feeder network.**

**Figure5. Gap (painted in black) in the area after introducing new feeder network.**

## 5. Conclusion

This study has focused on the expansion of the public transport network by filling gap areas. With a strongly connected PTN which covers the urban area, accessibility to the system will raise; subsequently, this will increase the interest in using the public transit system. In fact, this trend can potentially decrease travel time, cost of living in big cities, city costs, air pollution, fuel consumption, traffic jams and traffic accidents. Using feeder services is an economical and efficient solution to this problem. Therefore, designing an appropriate feeder network for gap areas can effectively develop the whole system. In this paper, we proposed a mathematical model, based on potential demand, to design feeder routes with various optional stations. In this model, different levels of interest for reaching each main station from different links have been considered, as well as we used labelling to ensure the connectivity between stations and routes. Then, we introduced a multi-parent crossover genetic algorithm to solve the model, in which initial parents are generated by a VRP algorithm. Tehran District 10 was selected as this paper's case study. The related database included some missing values in which they were estimated using a machine learning method along with considering seven features to obtain a more decent approximation. After generating the model by genetic algorithm, a four-route feeder network was obtained.

The accessibility rose from about 12% in the area without feeder routes to over 98% with the resulted network. Also, almost two-fifths of links have been connected to stations with an upper level of interest. The results show that the origin-destination travel time is momentous, and selecting a proper station and a railway line can cut off transfers number and travel times of commuters. In brief, considering all travel demands, including out of regional and intra-regional travels, should be studied for designing a multi-circulation feeder network.

For future work, it is intended to extend this model for designing the whole public transit network in a city and expand the model for design networks in disaster management.

# Tables

**Table1. Characteristics of obtained feeder routes.**

| Article | Objective | Decision Variables | Travel Demand Pattern | Optimization Method |
|---|---|---|---|---|
| (Kuah and Perl 1989) | Maximize the demand potential of the route links | Bus route and frequency | Based on demand nodes (bus stops) | A heuristic algorithm |
| (Martins and Pato 1998) | Minimize the total time | Bus route and frequency | Based on demand nodes (bus stops) | A combined building plus improving heuristic procedure, |
| (NEE 2004) | Minimize the total time | Bus route and frequency | Based on demand nodes (bus stops) | A heuristic algorithm |
| (I-Jy Chien 2005) | Minimize the total cost | Bus route and fleets | Based on demand nodes (bus stops) | An integrated method considering both analytical and numerical techniques |
| (Ceder 2016) | Maximize the demand potential of the route links | Bus route | Based on route links | |
| (Chandran and Raghavan 2008) | Minimize the total distance | Bus route | Based on demand nodes (bus stops) | A heuristic algorithm |
| (Li, Lam and Wong 2009) | Minimize the total cost | Bus route and fleets | Based on demand nodes (bus stops) | A heuristic algorithm |
| (Li and Quadrifoglio 2011) | Minimize the total cost | Bus route | Based on residential zones | A heuristic algorithm |
| (Martínez and Eiró 2012) | Minimize the total distance | Bus route and fleets | Based on demand nodes (bus stops) | A heuristic algorithm |
| (Hu, Zhang and Wang 2012) | Minimize the total time | Bus route | | A represented algorithm |
| (Chandra and Quadrifoglio 2013) | Minimize the distance | Bus route | Based on demand nodes (bus stops) | A heuristic procedure |
| (Deng et al. 2014) | Minimize the total passenger travel expense and maximize the total operating profit | Bus stop location and route | An elastic demand based on time period in an area | |
| (Zhu et al. 2017) | Maximize the summation of travel demand | Bus route | Based on route links | A genetic algorithm |
| (Tabassum et al. 2017) | Minimize weighted features (cost, time, comfort,...) | Bus route | Based on demand nodes (bus stops) | |
| (Chen et al. 2017) | Minimize the travel time | Bus route | Based on route links | A two-stage algorithm |
| (Almasi et al. 2018) | Minimize the total cost | Bus route and frequency | Based on demand nodes (bus stops) | A meta-heuristic approach |
| Ramos, de Morais and Barbosa-Póvoa (2018) | Minimize transport cost | Vehicle route | Based on demand nodes | |
| Park, Lee and Sohn (2019) | Maximize the cosine similarity | Bus route | Based on residential zones | A branch-and-bound algorithm |
| (Lee et al. 2019) | Minimize the travel time and cost | Passenger serving | | A meta-heuristic approach |
| (Huang et al. 2020) | Maximize the operator's revenue | Bus route and request-to-vehicle variable | Based on demand nodes (bus stops) | Heuristic methods |

**Table2. Characteristics of obtained feeder routes.**

| Feeder route | 1 | 2 | 3 |
|---|---|---|---|
| Length | 6 km | 6 km | 6 km |
| Number of covered stops | 6 | 9 | 6 |
| Summation of potential demands for each route ≈ | 1166940 | 1221951 | 765230 |

**Table3. Comparing results of four different samples solved by the algorithm.**

| Sample | The number of initial parents | The number of generations | Best answer | Running time (second) |
|---|---|---|---|---|
| Randomized 1 | 15 | 100000 | 1.3305e+07 | 50.17 |
| Randomized 2 | 10 | 100000 | 1.3663e+07 | 34.51 |
| Benchmark 1 | 30 | 100000 | 1.9099e+07 | 90.28 |
| Benchmark 2 | 15 | 100000 | 1.5130e+07 | 48.66 |

**Table4. Three levels of satisfied demand from links for reaching PTN stations in final feeder network.**

| The level of connection between each link and main station | Percentage of links connected to stations |
|---|---|
| Upper level | 40% |
| Middle level | 28% |
| Lower level | 32% |

# Figures

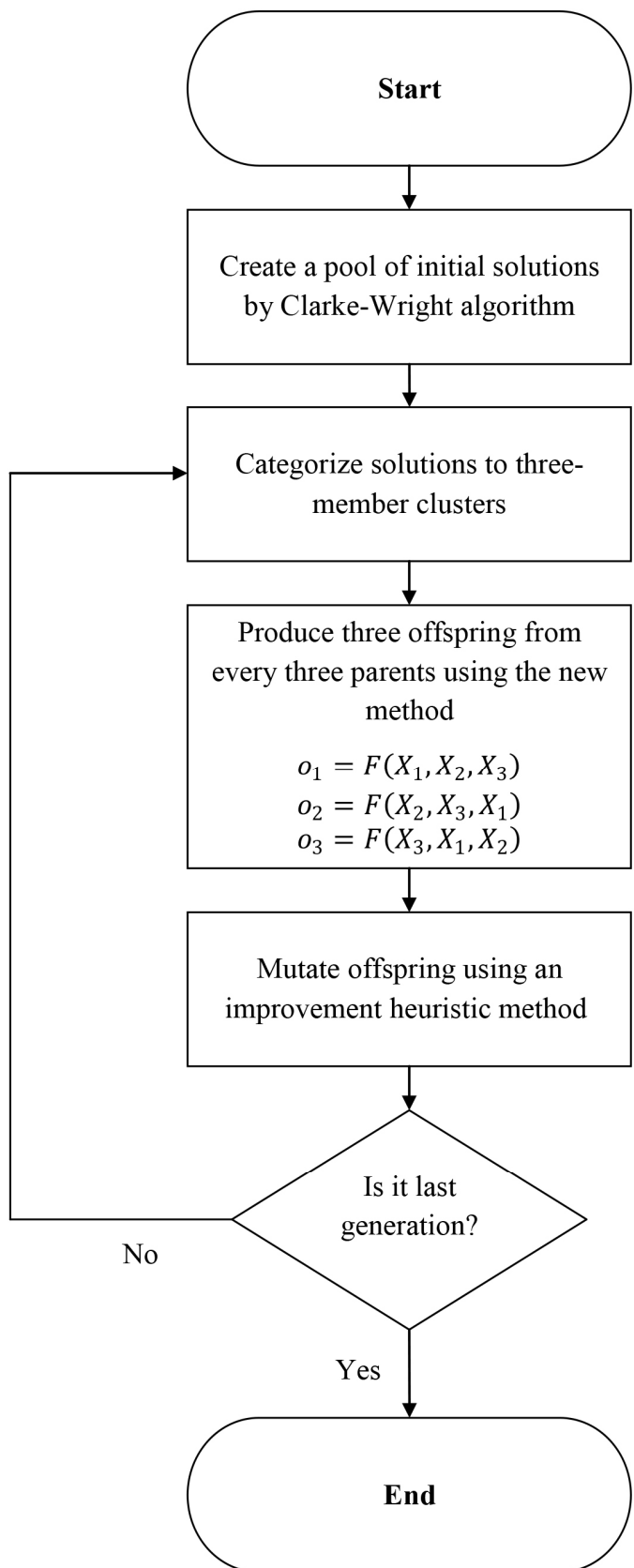

**Figure1. Flowchart for the proposed algorithm to solve the problem.**

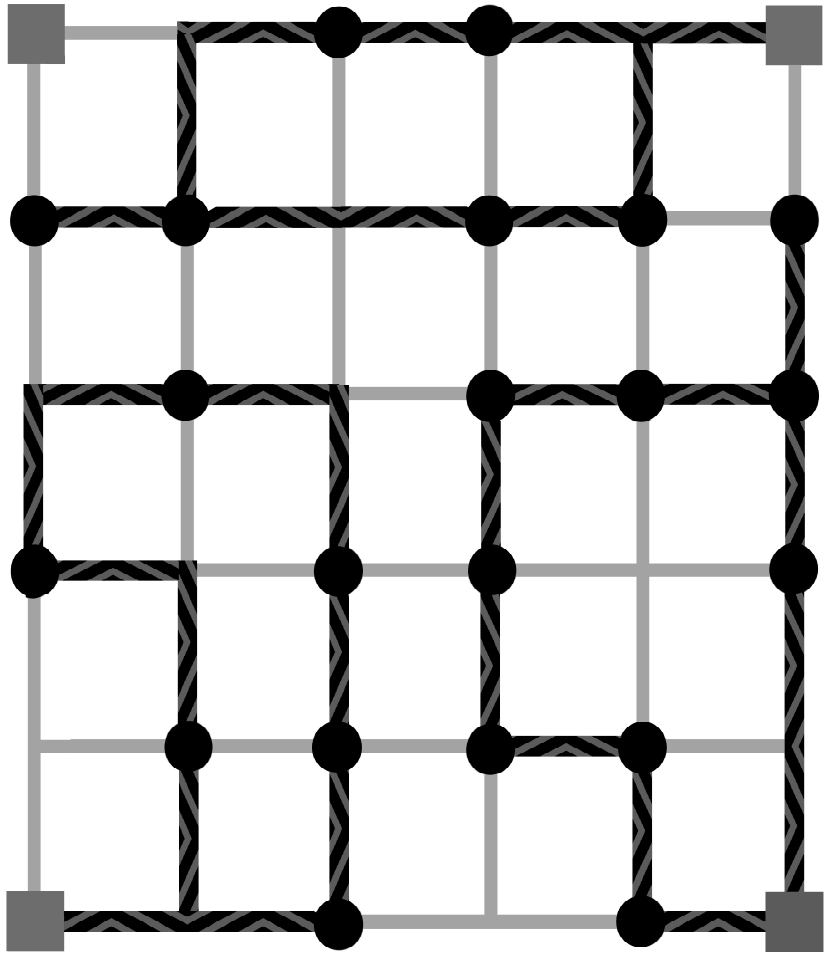

Main network station

Stop

Road

— Feeder route

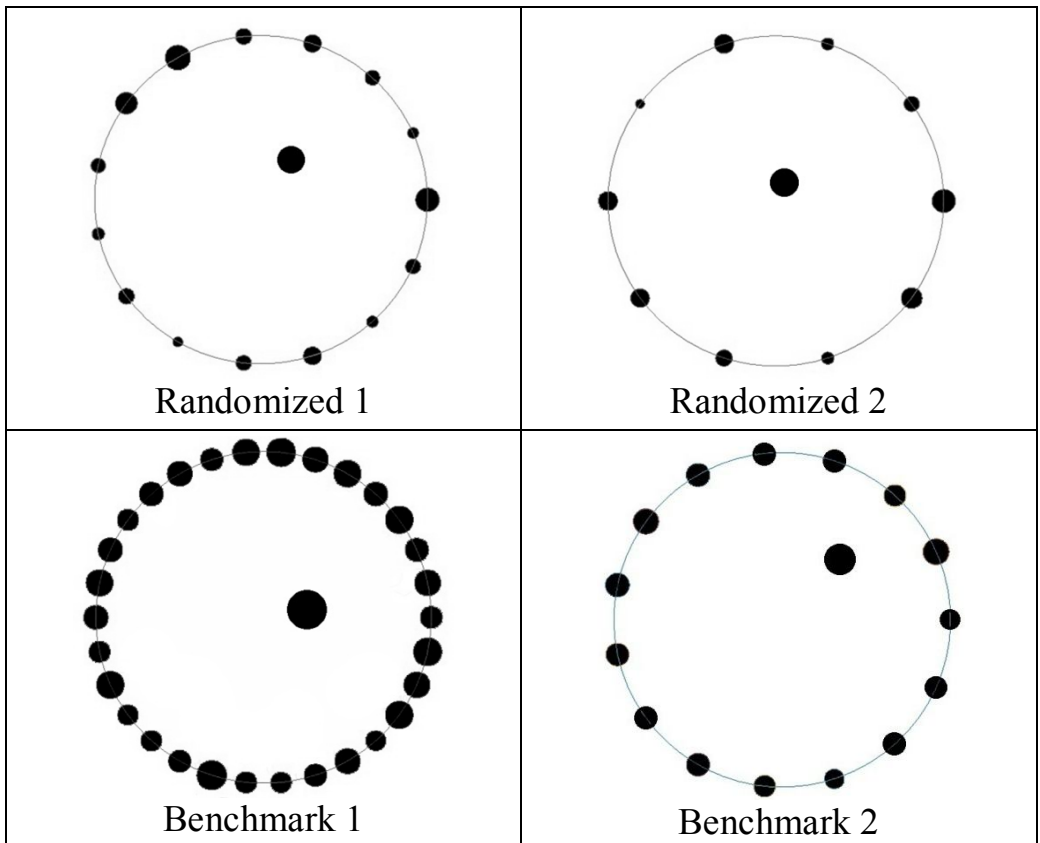
**Figure3. Resemblance between final answer and initial solutions for four samples.**

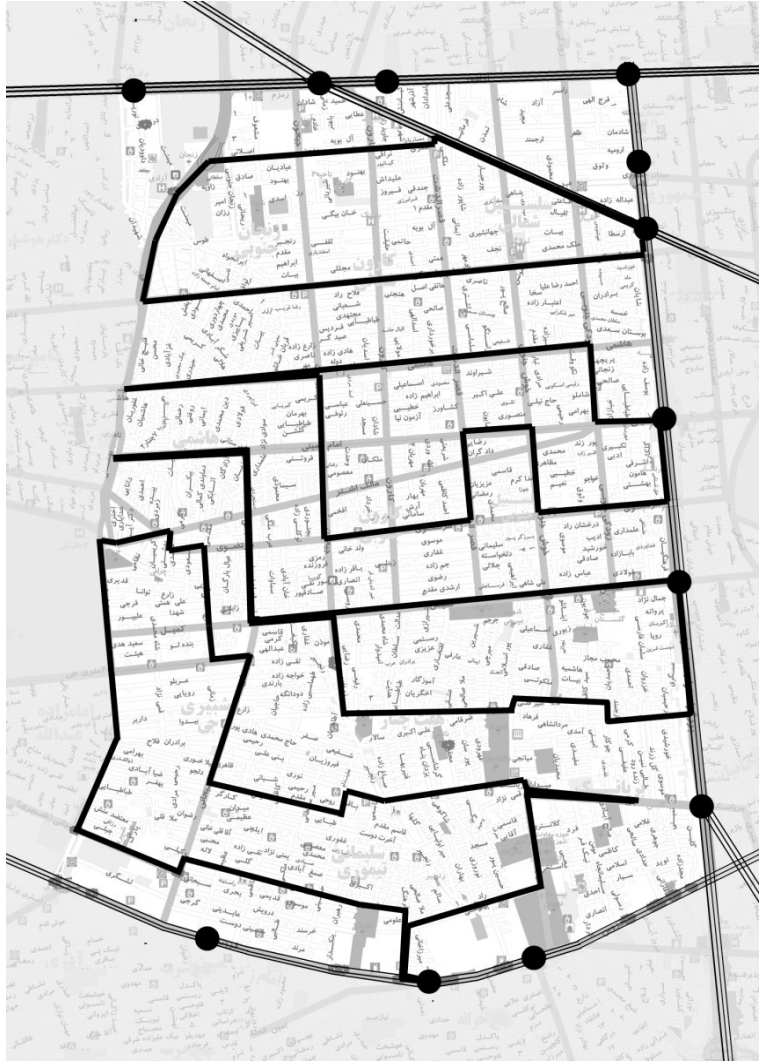

**Figure4. Proposed feeder network.**

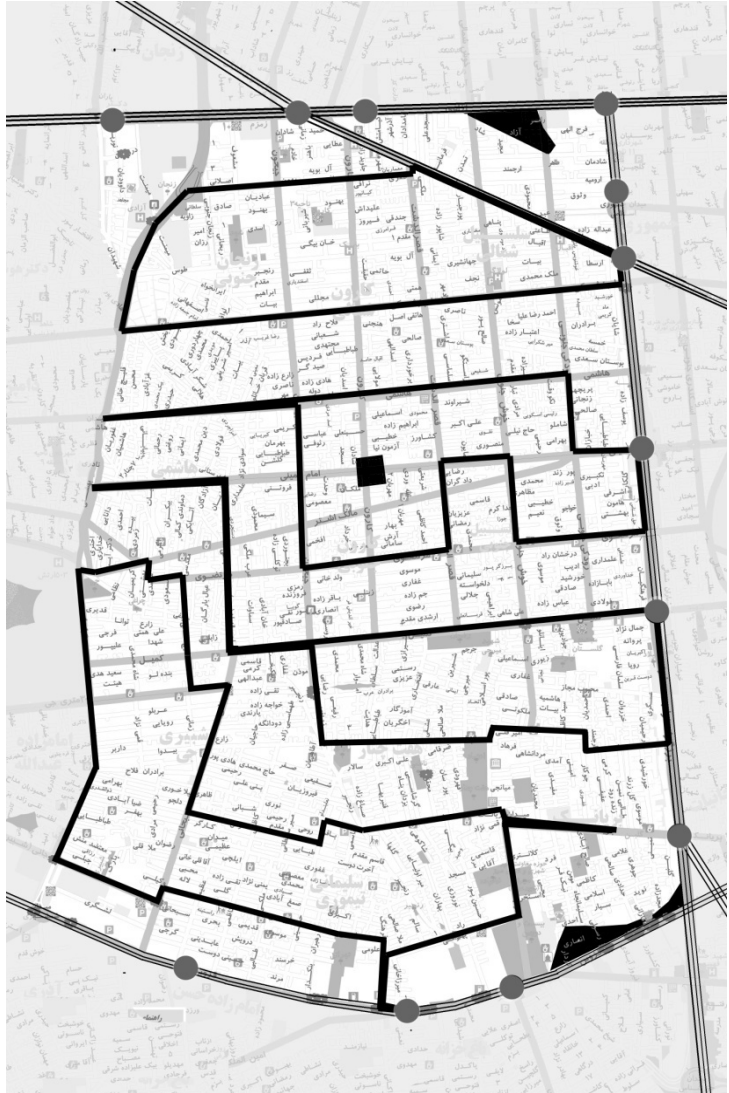

**Figure5. Gap (painted in black) in the area after introducing new feeder network.**